\documentclass[%
 reprint,
nofootinbib,
 amsmath,amssymb,
 aps,
pra,
]{revtex4-2}
\usepackage{hyperref}
\usepackage{titlesec}
\usepackage{caption}
\usepackage{subcaption}
\usepackage{graphicx}
\usepackage[export]{adjustbox}

\usepackage{flushend}
\usepackage{tabularx}
\usepackage{multirow}
\usepackage{cleveref}
\usepackage{dcolumn}
\usepackage{bm}
\usepackage{algorithm2e}
\usepackage{algorithmic}%
\usepackage{amssymb}
\usepackage{ulem}
\RestyleAlgo{ruled}
\usepackage{caption}
\usepackage{color}
\usepackage{appendix}

\begin{document}


\title{Overcoming the Standard Quantum Limit with Electro-Optomechanical Hybrid System for Enhanced Force  Sensing}
\author{Alolika Roy $^1$, Amarendra K. Sarma $^1$}
{ \affiliation{ 
    $^1$ Department of Physics, Indian Institute of Technology Guwahati, Guwahati 781039, India.}

\date{\today}

\begin{abstract}
We investigate the reduction of measurement-added noise in force sensing by analysing its power spectral density (PSD) within a hybrid optomechanical system. The setup comprises of an optomechanical cavity equipped with a movable mirror which acts as the mechanical oscillator, a stationary semi-transparent mirror, a superconducting qubit, and an optical parametric amplifier (OPA). By utilizing the concept of coherent quantum noise cancellation (CQNC), we derive the conditions necessary for complete cancellation of back-action force, thereby enhancing force sensitivity. Furthermore, with the gradual increase in the OPA pump gains, we suppress the sensitivity beyond the standard quantum limit (SQL) at a lower value of laser power. The removal of back-action noise, along with the reduction of shot noise, improves force detection capabilities, thereby surpassing the standard quantum limit associated with weak force detection.

\end{abstract}

\maketitle

\section{Introduction}

Optomechanical systems have seen widespread adoption across various fields, including quantum information processing and communication \cite{stannigel2011optomechanical,stannigel2010optomechanical}, quantum correlations \cite{bemani2019quantum}, and squeezing \cite{dalafi2018effects}. Recognized as a powerful tool for high-precision sensing \cite{caves1980measurement,danilishin2012quantum}, optomechanics \cite{RevModPhys.86.1391,clerk2010introduction, chen2013macroscopic} has proven particularly effective in applications such as measurement of acceleration \cite{krause2012high, qvarfort2018gravimetry}, mass \cite{liu2019realization,bin2019mass}, acoustic signals \cite{monifi2013ultrasound,chistiakova2014photoelastic,yang2020multiphysical}, displacement \cite{rossi2018measurement,matsumoto2019demonstration,mason2019continuous}, and weak force detection \cite{caves1980measurement,schreppler2014optically,zhou2020spectrometric,PhysRevLett.121.031101}. Typically, an analysis of the noise spectral density is performed to estimate the optimal operating frequencies of these systems and thereby, enhance their sensitivities.



Noise can arise from various sources, including thermal noise caused by environmental factors and measurement-induced noise of quantum mechanical origin. In precision measurement the quantum mechanical noises, namely shot noise and back-action noise play a crucial role  in determining the sensitivity limit. The interplay between the two noises leads to optimum balanced point known as the standard quantum limit (SQL) \cite{bowen2015quantum,caves1980measurement,clerk2010introduction,danilishin2012quantum}, which sets a lower bound of sensitivity. 
In standard optomechanical system, force sensitivity cannot be achieved beyond the SQL as back-action noise and shot noise demonstrate opposite responses to the laser power. If we can eliminate the back-action noise completely with some arrangements, then we will be able to
regulate the laser power to reduce the shot noise further.
This will lead to a significant reduction in measurement
added noise.

Various techniques have been introduced to reduce the effect of measurement added noises such as variational measurement \cite{kimble2001conversion}, introducing Kerr medium \cite{bondurant1986reduction}, squeezing the input light \cite{bondurant1984squeezed,chelkowski2005experimental, mow2004experimental}, optical spring effect \cite{chen2011qnd} etc. Quantum noise cancellation is a potential method to improve the performance of force sensors beyond the SQL. Back-action evasion by Coherent Quantum Noise Cancellation (CQNC) has gained significant attention over past decade. This scheme has been proposed for the first time by Tsang and Caves \cite{tsang2010coherent} by adding an anti-noise path in an auxiliary cavity coupled to the main optical mode of an optomechanical system. This was to ensure that the response of the auxiliary cavity for the amplitude fluctuations cancel out the noise in the primary cavity, thereby eliminating the back-action noise.


 Later, some other systems also reported noise cancellation which include inverted atomic spins \cite{hammerer2009establishing} and negative mass BE condensates \cite{zhang2013back}. 



Recently CQNC in hybrid optomechanical systems have also gathered attention, where the optical mode is coupled to atomic gas \cite{Motazedifard_2016,bariani2015atom,singh2023enhanced} and reported a significant noise reduction. Such systems are known to achieve improved optomechanical cooling \cite{PhysRevA.80.061803, PhysRevA.91.013416, PhysRevA.90.033838}. In the above investigations, the auxiliary system is realized by an atomic ensemble which is essentially assumed to act like a two level system. Additionally, the use of Optical Parametric Amplifier (OPA) further improves high precision detection as discussed elaborately in Refs. \cite{peano2015intracavity,singh2023enhanced}. Noise reduction in hybrid system has been observed experimentally as well \cite{moller2017quantum}.




We introduce a new approach of improving sensitivity with a hybrid electro-optomechanical(EOM) model, which comprises of optical and electrical components along with a mechanical resonator \cite{barzanjeh2011entangling,malossi2021sympathetic,PhysRevA.104.023509}. Here, the mechanical oscillator is coupled to both superconducting qubit and optical cavity \cite{PhysRevA.104.023509,blais2021circuit,nongthombam2023synchronization}. The electromechanical couplings have been demonstrated theoretically \cite{martin2004ground,vitali2007entangling,kounalakis2020flux} as well as experimentally \cite{lahaye2009nanomechanical,chu2017quantum}. While the Ref. \cite{PhysRevA.104.023509} explores the hybrid EOM system in ground state cooling, such type of system has not been explored in the context of precision sensing. However, it can be identified as a potential platform for improved sensitivity. Our study presents an approach to reduce noise by back-action force cancellation technique with this model. Furthermore, utilizing OPA force sensing can significantly be improved to reach sensitivity beyond SQL. This approach promises noise reduction by several orders of magnitude at off-resonant frequency.



 The paper is organized as follows: In Section II, we introduce the system and describe its Hamiltonian. From this Hamiltonian, we derive the quantum Langevin equations of motion, which provide the dynamics of the system. In Section III, after solving the Langevin equations, we obtain the output phase quadrature and address the force sensing and noise cancellation scheme. Section IV discusses the results, and finally, we conclude by summarizing the work in Section V.


\section{The System}
 Fig. 1 illustrates our hybrid electro-optomechanical model \ref{cavity_Hybrid} which comprises of an optical cavity of resonance frequency $\omega_c$, a mechanical oscillator of frequency $\Omega$ and a superconducting qubit \cite{PhysRevA.104.023509}. The mechanical mode couples to the optical mode via radiation pressure and to a Cooper-pair box (CPB)
qubit via a movable capacitive plate with capacitance $C_x(x)$. The optomechanical cavity is driven by a laser field of frequency $\omega_L$ and input power, $P_L$. The schematic diagram is shown in Fig.1.

\begin{figure}[!ht]
\centering
      \includegraphics[width=.5\textwidth]{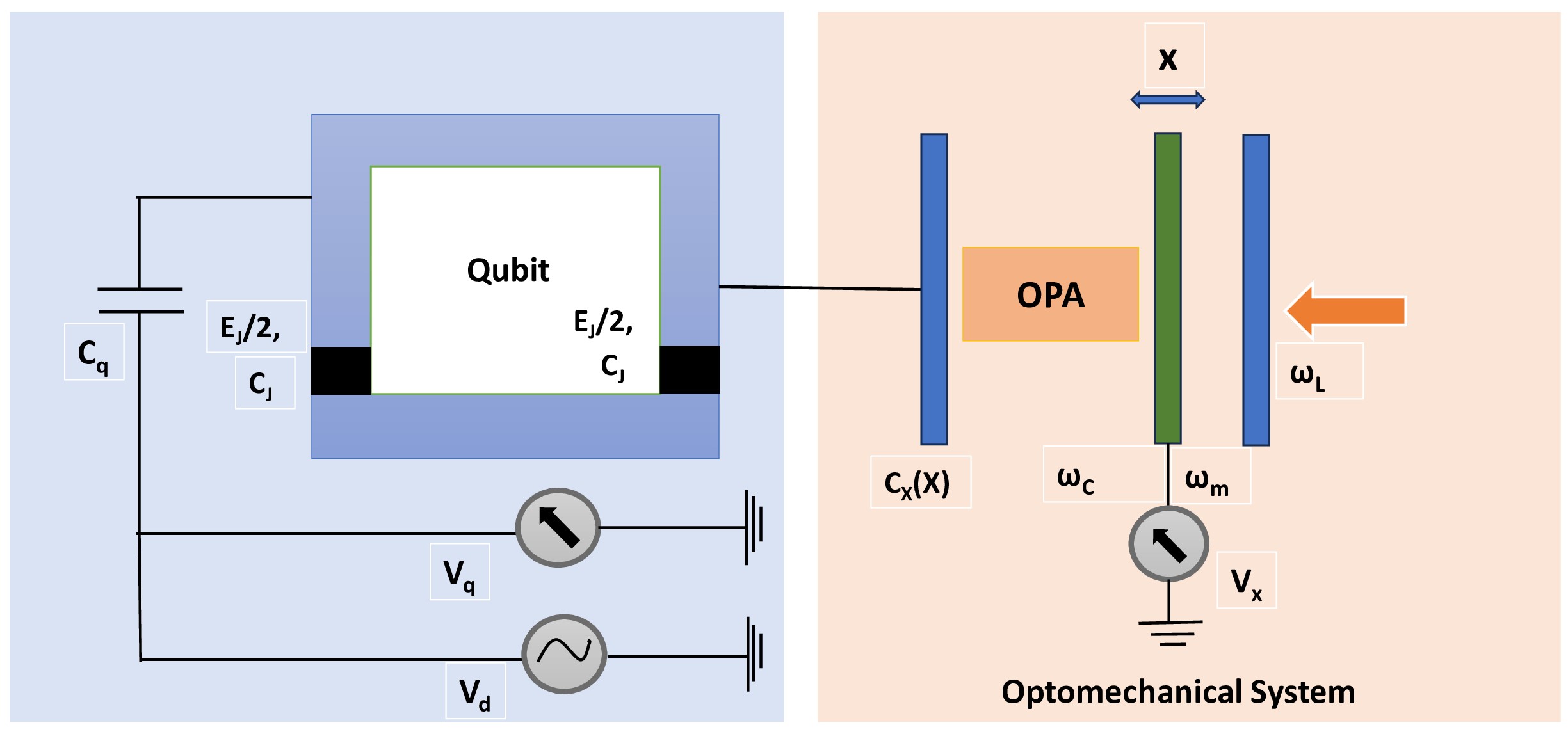}
      \caption{Schematic diagram of a Hybrid Electro-optomechanical system. The CPB is biased by voltages $V_q$ and $V_x$, and driven by voltage, $V_d$. The total capacitance of the qubit is $C_{\Sigma} (x) = 2C_J + C_q +C_x(x)$. 
      \label{cavity_Hybrid}}
\end{figure}
This system Hamiltonian can be expressed as follows:
\begin{align}
    H=H_{OM} +H_{OPA}+H_{L} +H_{Q},
    \label{Hamiltonian}
\end{align}

 where $H_{OM} = \hbar \Omega \hat{b} ^{\dagger} \hat{b} - \hbar \Delta _c \hat{a} ^{\dagger} \hat{a} + \hbar g_0  \hat{a} ^{\dagger} \hat{a} (\hat{b} ^{\dagger} 
 + \hat{b})$ describes the Hamiltonian of the standard optomechanical system. The first part of \( H_{OM} \) represents the contribution from the mechanical mode, the second term corresponds to the optical mode, and the final part arises from the optomechanical interaction with coupling strength \( g_0 \). Here, \( \Delta_c \) denotes the detuning between the laser drive and the optical mode frequency, defined as \( \Delta_c = \omega_L - \omega_c \). The operators \( \hat{a} \) (\( \hat{a}^\dagger \)) and \( \hat{b} \) (\( \hat{b}^\dagger \)) are the annihilation (creation) operators for the cavity mode and the mechanical mode, respectively. 
 
 The next term \( H_{OPA} = i\hbar \mathcal{G} (\hat{a}^{{\dagger} 2} e^{i \theta} - \hat{a}^2 e^{i \theta}) \), represents the OPA coupled to the system \cite{singh2023enhanced}. Here, \( \mathcal{G} \) is the gain of the OPA, and \( \theta \) is the phase of the OPA pump. We can set the phase to zero, simplifying the Hamiltonian to \( H_{OPA} = i\hbar \mathcal{G} (\hat{a}^{{\dagger} 2} - \hat{a}^2) \). 
 
 The energy of the driving laser with amplitude \( E_L \) is given by \( H_L = i\hbar E_L (\hat{a}^\dagger - \hat{a}) \).  Here $E_L = \sqrt({P_L \kappa}/ \hbar \omega_L)$. 
 
 Finally, $H_Q = -\hbar \frac{\Delta _q}{2} \sigma _z + \frac{1}{2} \hbar \Omega _ R \sigma_x + \hbar G(\hat{b} ^\dagger + \hat{b}) \sigma_z$ represents the electromechanical contribution (see Appendix \ref{Hamiltonian_Append}) in the drive frame of the qubit, under the rotating wave approximation. Here, the qubit is driven coherently with amplitude $\Omega_R$. The detuning,  $\Delta_q = \omega_L - \omega_q$, with $\omega_q$ denoting the transition frequency of the qubit. The first two terms of ${H_Q}$ indicate the qubit energy. The last term accounts for the qubit-phonon interaction of strength $G$. The operator $\sigma_x$ and $\sigma_z$ are the Pauli matrices corresponding to the qubit. The Pauli matrices $\sigma_x$, $\sigma_z$, $\sigma^+$ and $\sigma^-$ satisfy the following relations, $[\sigma_z,\sigma^+]=2\sigma^+$, $[\sigma_z,\sigma^-]= - 2\sigma^-$ and $[\sigma^-,\sigma^+]=\sigma_z$ , where $\sigma_x = \sigma^+ + \sigma^-$. In the limit of low excitation of the qubit \cite{sete2012controllable, chang2011multistability, genes2008emergence, LiChen+2020+14+23} where $\sigma_z \simeq \langle\sigma_z\rangle \simeq -1$ and hence  $\sigma^+$ and $\sigma^-$ satisfy the bosonic commutation relation $[\sigma^-,\sigma^+]=1$. In this case the bosonic annihilation  (creation) operator $\hat{d}(\hat{d}^\dagger)$ can be considered corresponding to the two level system (qubit) and the final Hamiltonian takes the form,
\begin{align}
    H=&\Delta _q\hat d^\dagger\hat d +\frac{1}{2} \Omega_R (\hat d^\dagger +\hat d)
    +\frac{\hbar \Omega}{2}(\hat x^2+\hat p^2)\notag \\ 
    &-\hbar {\Delta_c} \hat a^\dagger \hat a
     + \hbar g_0 \alpha ( \hat a^\dagger+ \hat a)\hat X  + i\hbar \mathcal{G}( \hat a^{\dagger 2} - \hat a^2) \notag \\
    &+G \bar d\hat X (\hat d^\dagger +\hat d)+ 2 G\bar X \hat d^\dagger \hat d + i\hbar E_L (a^{\dagger} - a)
\end{align}
Here, $\hat{x}$ and $\hat{p}$ are the dimensionless position and momentum quadrature respectively \cite{RevModPhys.86.1391}.
  In the strongly driven cavity field and in weak optomechanical coupling limit of the qubit, the system dynamics can be expressed as linearized quantum Langevin equation. 
 In the linearized approach, a generic operators can be written as quantum fluctuation around their respective  steady state as, $\hat{A} = \bar{A} + {\delta {\hat{A}}}$,  where $\bar{A}$ is the mean field value and $\delta \hat{A}$ is small quantum fluctuation. The higher orders of quantum fluctuations can be omitted. The  amplitude and phase quadratures of the optical, mechanical and qubit modes can be  defined as,  $\hat{x}_A =\frac{(\hat{A} + \hat{A} ^\dagger)}{\sqrt{2}}$  and $\hat{p}_A = i\frac{(\hat {A}^\dagger - \hat {A})}{\sqrt{2}}$ respectively where $\hat{A}= \hat{a},\hat{b} ~\rm{and} ~\hat{d}$. $ A^{in} = \hat{a}^{in}, \hat{b}^{in} ~\rm{and} ~ \hat{d}^{in}$  indicate the noise operators associated with the three modes. They satisfy the relations \cite{walls1994gj, benguria1981quantum,gardiner1985input, wimmer2014coherent} $\langle\hat{a} _{in} (t) \hat{a} ^\dagger_{in}(t^\prime) \rangle = \delta (t- t^ \prime)$ , $\langle\hat{b} _{in} (t) \hat{b} ^\dagger_{in}(t^\prime) \rangle = \delta (t- t^ \prime)$ and $\langle\hat{d} _{in} (t) \hat{d} ^\dagger_{in}(t^\prime) \rangle = \delta (t- t^ \prime)$. The corresponding amplitude and phase of noise terms are introduced as $\hat{X}^{in}_A =\frac{({\hat{A}}^{in} + \hat{A} ^{{in}\dagger})}{\sqrt{2}}$  and $\hat{P}^{in}_A = i\frac{(\hat {A}^{{in}\dagger} - \hat {A}^{in})}{\sqrt{2}}$.
     
The linearized quantum Langevin equations are given by, 
\begin{align}
     &\dot {\hat x}_a=(-\frac{k}{2}+2\mathcal{G}) x_a +\sqrt{\kappa} x^{in}_a \\
     &\dot {\hat p}_a=-(\frac{\kappa}{2}+2\mathcal{G})\hat p_a-\sqrt{2}g_0\alpha \hat x +\sqrt{\kappa }p^{in}_\alpha\\
     &\dot{\hat x}=\Omega \hat p\\
     &\dot {\hat p}=-\Omega \hat x -\sqrt{2}g_0 \alpha \hat x_a-\gamma_m \hat p+\sqrt{\gamma_m}(f_{th}+f_{ext}) \notag\\
     & ~~ ~~ -\sqrt{2}G \bar d x_d\\
     &\dot{\hat x}_d=\Delta_q\hat p_d + 2 G \bar x \hat p_d-\frac{\Gamma}{2} \hat x_d +\sqrt{\Gamma} x^{in}_d\\
     &\dot{\hat p}_d=-\Delta_q x_d-2\sqrt{2}G\bar d \hat x -2 G\bar x\hat x_d  +2\Omega _R \bar d \hat x_d\notag\\
     & ~~ ~~ ~  -\frac{\Gamma}{\sqrt{2}}\hat p_d+\sqrt{\Gamma}p_d^{in}.
     \end{align}
Here, $\bar x, ~\bar d$ are the steady state values. $f_{th}$ is the Brownian type thermal noise and $f_{ext}$ is the external force acted on the mechanical oscillator. The mechanical damping, cavity decay and qubit dephasing rate are represented by $\gamma_m$, $\kappa$ and $\Gamma$ respectively.

The linear response \cite{RevModPhys.86.1391} of the system can be studied by transforming the dynamics of the system from the time domain to the frequency domain. The Fourier transformation defined by $O(\omega) = \frac{1}{\sqrt{2\pi}} \int dt O(t) e^{- i\omega t} $ can be utilized in writing the Eqs. (3)-(8) in frequency domain as follows:

\begin{align}
    &\hat X (\omega)=\chi_m\{-g\hat X_a(\omega)+\sqrt{\gamma_m}(F_{th}(\omega)\notag\\
     & ~~ ~~ ~~ ~~ ~~+F_{ext}(\omega))-G^\prime \hat X_d (\omega)\}\\
     &
     \hat P(\omega)=\frac{i \omega}{\Omega} \hat X(\omega)\\
     &
     \hat X_a(\omega)=\sqrt{\kappa}\lambda_+X^{in}_a(\omega)\\
     & \hat P_a(\omega)=g^2 \chi_m \lambda_-\lambda_+\sqrt{\kappa}X^{in}_a(\omega)\notag\\
     &~~~~~~~~ -g{\lambda_-}\chi_m \sqrt{\gamma_m}(F_{th}+F_{ext}) +g G^\prime \chi_m \lambda_-\hat X_d(\omega) \notag\\
     &~~~~~~~~ +\lambda_- \sqrt{\kappa} P^{in}_a\\
     &     \hat X_d(\omega)=\Delta_q \chi_d \hat P_d (\omega)+ \chi_d \sqrt{\Gamma}x^{in}_d\omega)\notag\\
     &\hat P_d(\omega)=\zeta \chi_d \Omega^\prime \sqrt{\Gamma} \hat X_d^{in}+ 
     2\zeta g G^\prime \chi_m \sqrt{\kappa}\lambda_+X^{in}_a \notag\\
     &~~~~~~~~ -2 \zeta G^\prime\sqrt{\gamma}\chi_m[F_{th}(\omega)
      +F_{ext}(\omega)] \notag\\
      &~~~~~~~~ +\zeta G^{\prime 2} \chi_m \chi_d \sqrt{\Gamma}X^{in}_d(\omega)+\zeta \sqrt{\Gamma}\hat P^{in}_d.
 \end{align}

 Here, $\chi_d= \frac{1}{i\omega+\Gamma/2}$, $\chi_m=\frac{\Omega}{\Omega^2-\omega^2+i\gamma_m\omega}$, $\zeta = \frac{1}{i\omega + \Gamma /2 + \Delta_q ^2 \chi_d}$, $G^{\prime} = \sqrt{2} G \bar d$, $g= \sqrt{2}g_0 \alpha $ $\lambda_+ = \frac{1}{\chi^{-1}_a - 2 \mathcal{G}}$ and $\lambda_- =\frac{1}{\chi^{-1}_a + 2 \mathcal{G}}$, with $\chi_a = \frac{1}{i\omega + \kappa /2}$ . The scaled external force and thermal noise are defined as, $F_{ext} = \frac{f_{ext}}{\sqrt{\hbar m \omega _m \gamma _m}} $ and $F_{th} = \frac{f_{th}}{\sqrt{\hbar m \omega _m \gamma _m}}$ respectively \cite{allahverdi2022homodyne}. The scaled thermal Brownian force satisfies the relations $\langle F_T (t)F_T (t^\prime) \rangle = \bar {n} \delta (t-t^\prime)$, where $\bar n = \frac{K_B T}{\hbar \Omega}$ \cite{wimmer2014coherent}.

\section{Force sensing with coherent quantum noise cancellation}
\label{CQNC}
When the external force ($F_{ext}$) acts on the movable mirror, its position and consequently the cavity length changes. This affects the phase of the output cavity field. We solve Langevin equations in the frequency domain to obtain the phase quadrature of cavity field ($P_a$), which is related to the output field via the standard input-output relation, $P^{out}_a = \sqrt{k}P_a - P^{in}_a$. 
This can be expressed as,

\begin{align}
      P^{out}_a=\notag& \lambda_+ \lambda_-  \kappa (g^2 \chi_m + G^{\prime 2}\chi_d\prime)x^{in}_a\\ 
      \notag
      &+\sqrt{\kappa}(-\Omega\chi_d\chi^\prime_d \sqrt{\Gamma} \lambda_-+\chi_d \sqrt{\Gamma}g G^\prime \chi_m \lambda_-)x^{in}_d\\
      \notag
      &+ \sqrt{\kappa}(-\Omega \chi_d)(g G^\prime \chi_m \lambda_-)(\sqrt{\Gamma}\zeta) p^{in}_d\\
      \notag
     &+(-g \chi_m \lambda_- \sqrt{\gamma_m \kappa})(F_{ext}+F_{th})\\
      \notag
     &+ (\lambda_-\kappa-1) p^{in}_a,\\
     &
      \label{paout}
\end{align}

where, 
\begin{align}
   &\chi ^\prime_d= -\Omega\chi_d\zeta = \frac{\Omega}{\Omega ^2 - \omega^2 + i \omega \Gamma + \Gamma ^2/4}                            \\
   &\chi_d=  \frac{1}{i\omega+\Gamma/2}                                    \\
   &\chi_m=\frac{\Omega}{\Omega^2-\omega^2+i\gamma_m\omega}
\end{align}

The first term of $P^{out}_a$ corresponds to the back-action noise \cite{allahverdi2022homodyne,Motazedifard_2016}. For significant noise reduction, we aim to eliminate the back-action noise completely \cite{tsang2010coherent,wimmer2014coherent}. In Eq.\ref{paout} an extra noise term  ${G^\prime}^2 \chi^{\prime}_d$ appears which corresponds to the electromechanical sub-system of the hybrid system . Under suitable conditions this term cancels the optomechanical back-action term $g^2\chi_m$, resulting in complete elimination of back-action noise for the hybrid system.
The back-action noise cancellation condition is 
 
  \begin{align}
   g^2\chi_m +{G^\prime}^2 \chi^{\prime}_d=0  
   \label{CQNC_main}
 \end{align}

The details of the derivation of the above mentioned condition is given in the Appendix \ref{constrain_Append}.
To satisfy the Eq.\ref{CQNC_main} we can choose the coupling and susceptibilities of the optomechanical and electromechanical sub-systems in such a way that $g=$  $G^\prime$ and $\chi_m=$$-\chi^{\prime}_d$.
 This implies that i) the response of mechanical oscillator to the back-action noise is exactly equal and opposite to that of superconducting qubit involved in the system and ii) the optomechanical coupling is equal to the electro-mechanical coupling of the hybrid system.

 Complete cancellation of backaction noise would allow us to overcome the SQL. In absence of back-action noise, the laser power can be increased resulting in further reduction in shot noise and consequently suppression of the total measurement added noise, which can be expressed in terms of the output phase quadrature.  
Furthermore, analysing the definitions of susceptibilities of the mechanical oscillator ($\chi_m$) and that of qubit ($\chi_d ^\prime$) we can further infer that they would cancel each other if ,
\begin{enumerate}

   \item the qubit is detuned from the cavity by $\Delta _q =  \Omega$,

\item the qubit dephasing rate matches the mechanical oscillator damping rate $\Gamma = \gamma _m$

\item  $|\Delta _q| \gg \Gamma$. The condition 1. would eventually give, $\Omega \gg \Gamma$. 

\end{enumerate}

These are the requirements for the perfect noise cancellation to be attained experimentally.
 Once we consider the noise cancellation condition given in eq. \ref{CQNC_main}, the output phase quadrature takes the following form,
 
\begin{align}
 P^{out}_{a}= 
      \notag
      &+\sqrt{\kappa}(-\Omega\chi_d\chi^\prime_d \sqrt{\Gamma} \lambda_-+\chi_d \sqrt{\Gamma}g G^\prime \chi_m \lambda_-)x^{in}_d\\
      \notag
      &+ \sqrt{\kappa}(-\Omega \chi_d)(g G^\prime \chi_m \lambda_-)(\sqrt{\Gamma}\zeta) p^{in}_d\\
      \notag
     &+(-g \chi_m \lambda_- \sqrt{\gamma_m \kappa})(F_{ext}+F_{th})\\
      \notag
     &+ (\lambda_-\kappa-1) p^{in}_a\\
     &
      \label{paout_cancelledBA}    
 \end{align}

 We can re-write the above expression as,
 \begin{align}
    &\frac{P^{out}_{a}}{(-g\chi_m\lambda_-\sqrt{\gamma_m \kappa})} =  
    \notag
    F_{ext}+ \frac{(\lambda_-\kappa-1) p^{in}_a}{(-g\chi_m\lambda_-\sqrt{\gamma_m \kappa})} \\
    \notag
    &+ \frac{\sqrt{\kappa}(-\Omega\chi_d\chi^\prime_d \sqrt{\Gamma} \lambda_-+\chi_d \sqrt{\Gamma}g G^\prime \chi_m \lambda_-)x^{in}_d}{(-g\chi_m\lambda_-\sqrt{\gamma_m \kappa})}\\
    \notag
    &+\frac{\sqrt{\kappa}(-\Omega \chi_d)(g G^\prime \chi_m \lambda_-)(\sqrt{\Gamma}\zeta) p^{in}_d}{(-g\chi_m\lambda_-\sqrt{\gamma_m \kappa})}\\
    &
\label{paout_F_simplified}
\end{align}

We estimate the impressed force on the mechanical oscillator from the output phase quadrature. Based on Eq.\ref{paout_cancelledBA}, the force estimator can be written as, 
 
 \begin{align}
     \hat{F} = \frac{P^{out}_a}{(-g\chi_m\lambda_-\sqrt{\gamma_m\kappa})}=F_{ext}+F_{add}
 \end{align}
where, the  $F_{add}$ is the added noise given as,
 \begin{align}
     F_{add}=\notag
     & F_{th}-{\frac{\kappa\lambda_- - 1}{g\chi_m \lambda_-\sqrt{\gamma_m}\kappa} } P^{in}_a +  \frac{\sqrt{\Gamma}}{\gamma_m}P^{in}_d\\
     &-(\frac{i\omega+\Gamma/2}{\Omega})x^{in}_d
 \end{align}

\section{Results and Discussion}

We now proceed to the numerical analysis of the power spectral density which provides a measure of the force sensitivity \cite{Motazedifard_2016}. The spectral density can be defined as in \cite{wimmer2014coherent},
 \begin{align}
 S^{add}_F(\omega)\delta(\omega-\omega^\prime)=\frac{1}{2} (\langle\hat F_{add}(\omega)\hat F_{add}(-\omega)\rangle +c.c.)
 \label{S_F}
\end{align}

Using this we obtain the expression of spectral density of the added noise, $S_{F}^{add}(\omega)$ expressed as

\begin{align}
   S_F^{add} =
   \notag
   & \frac{K_B T}{\hbar \Omega}  + \frac{1}{2} \frac{1}{g^2 |\chi_m(\omega)|^2(\gamma_m \kappa)} (\frac{\lambda_-(\omega)\kappa - 1}{\lambda_-(\omega)})^2 \\
   &+ \frac{1}{2}\frac{\omega ^2 +\Omega ^2 + \Gamma ^2/4}{\Omega^2}
   \label{SF_added_1}
\end{align}

In order to study the measurement added noise reduction, the Brownian thermal noise background can be avoided\cite{Motazedifard_2016}. By minimizing the spectral density, we obtain the minimum noise spectral density for our proposed scheme in hybrid electro-optomechanical model as,

\begin{align}
    S_{F,CQNC}=\frac{1}{2}\frac{\omega^2+\Omega^2 +\Gamma^2/4}{\Omega^2}
\label{SF_CQNC}
\end{align}

We compare the efficiency our scheme with the standard optomechanical system. The spectral density for a standard optomechanical system, can be written as \cite{RevModPhys.86.1391},
\begin{align}
  S_F(\omega)=  \frac{K_B T}{\hbar \Omega}+\frac{1}{2}\frac{\kappa}{\gamma_m} \frac{1}{g^2 |\chi|^2} \frac{1}{4}+ 4g^2 \frac{1}{\kappa\gamma_m}
\label{SF_SQL}
\end{align}
 The first term indicates the contribution of the thermal noise, second term represents shot noise and the third term corresponds to the back-action noise. Minimizing this with respect to the laser power, we can get the SQL value,
\begin{align}
    S_{F,SQL}=\frac{1}{\gamma_m |\chi_m|}
\label{SQL_min}
\end{align}
Here the thermal force is considered to be negligibly small. The dip of the noise spectral density occurs at an optomechanical coupling strength, $ g_{SQL} = \sqrt{\kappa}/(2\sqrt{| \chi _m |})  
   \label{g_{SQL}} $.

\begin{figure}[h!]
\centering
    \includegraphics[height=5cm,clip]{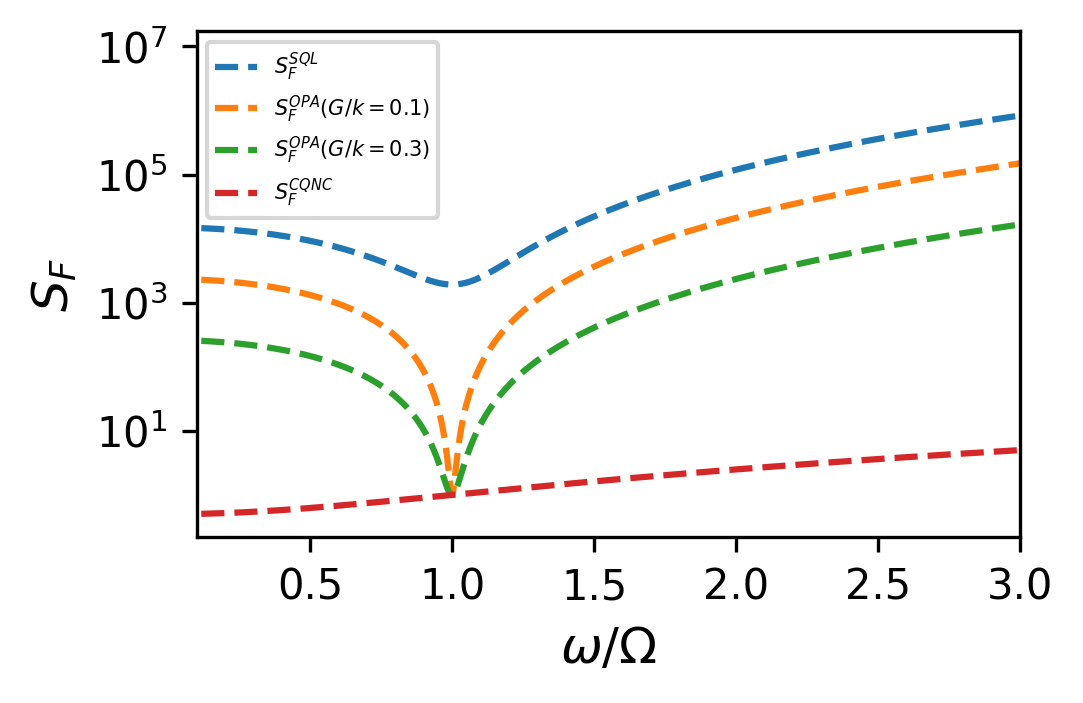}
\captionsetup{justification=raggedright}
\caption{Noise Power Spectral Density for standard optomechanical system, optomechanical system with OPA for parametric gain $G/\kappa = 0, 0.1$ and $0.3$ and optomechanical hybrid system with CQNC scheme. The spectral densities are normalized by $\hbar m \omega _m \gamma _m$ in order to be represented in units of $\rm{N^2 Hz^{-1}}$. The different lines in the plot represent the standard optomechanical system (blue curve), the hybrid electro-optomechanical system with OPA (orange and green curves), and the hybrid system with the CQNC scheme (the red line at the bottom). The parameters used \cite{wimmer2014coherent, singh2023enhanced}: $g_0 = 300 \times 2\pi$ Hz, $\Omega= 300 \times 2\pi$ KHz, $\gamma_m = 30 \times 2\pi$ Hz, $\kappa= 2\pi$ MHz, $P=100$ mW, $\omega_L = 384 \times 2\pi$ THz}.

\label{PSD1}
\end{figure}

\begin{figure}[h!]
\centering
    \includegraphics[height=5cm,clip]{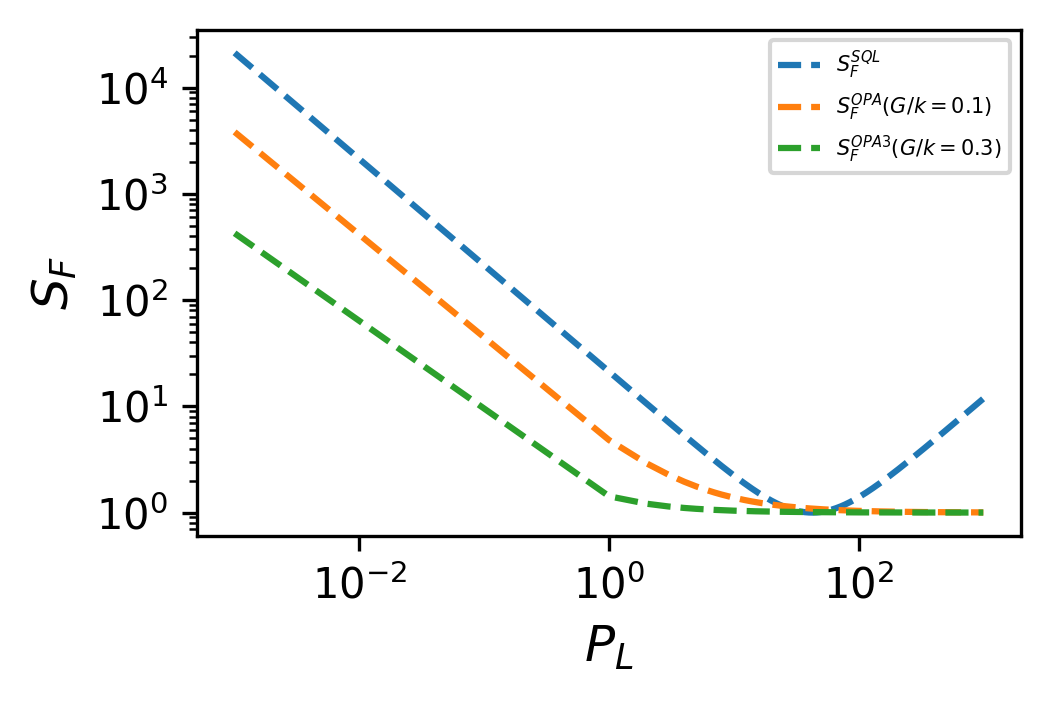}
  \captionsetup{justification=raggedright}

    \caption{Noise Power Spectral Density at resonance (i.e., $\omega=\Omega$) as a function of laser driving power for the standard optomechanical system and the electro-optomechanical hybrid system with the CQNC scheme. The blue line represents the standard optomechanical system, whereas, the orange and green curve indicate hybrid system with OPA gain $G/\kappa = 0.1$ and $0.3$ respectively. (Parameters are the same as in Fig.\ref{PSD1})}
\label{PSD_PL}
\end{figure}

In Fig. [\ref{PSD1}] we plot the variation of the noise spectral density of the optomechanical system as function of detection frequency.
When the system is at resonance, i.e. when the detection frequency $= \Omega$, we observe a dip in the spectral density ($S_F$) curve. Here, $S_F$  for the later two cases are nearly equal. However, if we increase the OPA gain $G$ from $0.1$ to $0.3$, the noise spectral density at off-resonant frequencies get reduced noticeably as compared to the standard optomechanical system, which is visible from the nature of the orange and green curve. Furthermore, in the case of CQNC, we observe that at frequencies both lower and higher than the resonance frequency, the noise PSD is suppressed by several orders of magnitude. Although the CQNC noise cancellation technique may not enhance sensitivity at the resonant frequency, it demonstrates a significant reduction in noise at off-resonant frequencies. 
The variation of noise PSDs vs laser driving power is shown in Fig. [\ref{PSD_PL}]. The laser driving power is proportional to the square of optomechanical coupling given by $P_L = 2\hbar {\omega _L} \kappa (g/g_0)^2$. The noise spectral density is expressed as a function of input laser power. 
The figure shows that in a standard optomechanical system (represented by the blue line), the noise spectral density initially decreases as laser power increases. However, after reaching the minimum point, it begins to rise again due to the significant back-action noise at higher laser power levels. Whereas, for the hybrid system (orange and green line corresponding to OPA gain 0.1 and 0.3 respectively) the PSD decreases with the laser power and after reaching the minimum, it does not increase any more. This happens because, when CQNC scheme is incorporated, it cancels out the back-action noise at higher laser driving power. 

Furthermore, with increasing OPA gain from 0.1 to 0.3, the minimum is achievable at a lower value of laser power.
This suggests that in the our framework, using an OPA and increasing its gain allows us to achieve the minimum spectral density at a lower value of \( P_L \).

\section{Conclusion}
Finally, to summarize, we aim for back-action noise cancellation and overall noise reduction for an optomechanical system. In such scheme a hybrid optomechanical system is often utilized, where the back-action noise arising from the optomechanical sub-system is cancelled by the response coming from the hybrid counterpart. In this paper, we study the noise reduction in an electro-optomechanical system. The superconducting qubit present in the circuit acts as a two level sub-system. This qubit being coupled to the mechanical oscillator, it is more easily engineerable from outside without disturbing the optical cavity arrangement. From the analysis, we observed that when the effect of the back-action noise on the electro-mechanical sub-system is equal and opposite to that of the optomechanical one, perfect evasion of back-action takes place and consequently improvement in sensitivity. Additionally we observe that with increasing OPA gain noise reduction is possible at a lower laser power for our framework. This scheme may be utilized in quantum information processing and communication, gravitational wave detection, quantum metrology and further cooling of the mechanical mode beyond what conventional methods allow.

\section*{Acknowledgements} We acknowledge insightful and valuable discussions with Sampreet Kalita regarding the work. A.R. gratefully acknowledges the support of a research fellowship from UGC, Government of India. A.K.S. acknowledges the grant from MoE, Government of India (Grant No. MoE-STARS/STARS-2/2023-0161).

\appendix
\section{Appendix A: Hamiltonian}
\label{Hamiltonian_Append}
The Hamiltonian of the system without dissipation is given by \cite{PhysRevA.104.023509},
\begin{align}
  \hat{H} = \notag 
  &\frac{[\hat{Q} - Q_{xq}(x)]^2}{2C_{\Sigma} (x)} - E_J(\Phi _{ext}) cos \theta + \hbar \Omega \hat{b} ^ \dagger \hat{b}\\
  &
  - \hbar \Delta \hat{a} ^ \dagger \hat{a} + \hbar g_0 \hat{a} ^ \dagger \hat{a} (\hat{b} ^ \dagger + \hat{b}) + \hbar \eta (\hat{a} ^ \dagger + \hat{a})
\end{align}
 where, $Q= 2e\hat{N}$ with $\hat{N}$ being the number operator for the transferred Cooper pairs, $Q_{xq}(x)$, is the gate charge produced by the external gate voltages $V_x$ and $V_q$. The total capacitance of the qubit is $C_{\Sigma} (x) = 2C_J + C_q +C_x(x)$. The effective energy in the two parallel junctions is $E_J (\phi _{ext}) cos \theta$ with energy $E_J/2$ each. Here, $\theta$ is the phase difference between the junctions and $\phi_{ext}$ is the external flux. 

 The Hamiltonian is then further simplified after making the transformation $\hat{a} = \alpha + \delta \hat{a}$ and $\hat{b} = \beta + \delta \hat{b}$ where, $\alpha$ and $\beta$ are the respective steady states of cavity and mechanical modes and $\delta \hat{a}$, $\delta \hat{b}$ are the fluctuations around the corresponding mean value. Subsequently the fluctuation variables are re-expressed as, $\hat{a} \implies \delta\hat{a}$ and $\hat{b} \implies \delta\hat{b}$. Also rotating wave approximation has been taken care of. The Hamiltonian takes the form \cite{PhysRevA.104.023509}, 
 \begin{align}
     \hat{H} = \notag
     &-\hbar \frac{\Delta_q}{2} \sigma _z + \frac{1}{2} \hbar \Omega_R \sigma_x + \hbar (G\hat{b}^\dagger + G^* \hat{b})\sigma _z
     +\hbar \Omega \hat{b}^\dagger \hat{b}\\
     \notag
     &- \hbar \Delta_c \hat{a}^\dagger \hat{a}+  \hbar (g_0 \alpha \hat{a}^\dagger + g_0 \alpha ^* \hat{a}) (\hat{b}^\dagger + \hat{b})\\
     &+H_a + H_b
 \end{align}
  where, $H_a = \hbar g_0 (\alpha ^* a + \alpha a^\dagger)(\beta + \beta ^*) + \hbar \eta (a + a^\dagger) - \hbar \Delta _c (\alpha ^* a + \alpha a^\dagger)$ and $H_b =\hbar g_0 |\alpha|^2(b + b ^\dagger) + \hbar \Omega (\beta b^\dagger + \beta ^* b)$
 
 Here, $\sigma_x$ and $\sigma_z$ are Pauli matrices. In the low excitation limit, $\sigma_z \simeq \langle\sigma_z\rangle \simeq -1$ and $\sigma^+$ and $\sigma^-$ satisfy the bosonic commutation relation $[\sigma^-,\sigma^+]=1$.The bosonic annihilation  (creation) operator $\hat{d}(\hat{d}^\dagger)$ are the corresponding to the two level system (qubit). Furthermore, following the same approach the qubit mode can be written as $\hat{d} = \bar d + \delta \hat{d}$.
 
\section{Appendix A: Derivation of the constraint}
\label{constrain_Append}

 The contribution to the back-action noise in the expression of $F_{add}$ comes from the term involving $x^{in}_a$. Solving the Langevin equation, we obtain the expressions for all the quadratures (X and P) out of which $P^{out}_a$ holds the most significance in our calculation. In the expression of $P^{out}_a$, the coefficient of $x^{in}_a$ is,
\begin{align}
  [x^{in}_a]_{Coef}= \notag 
  & \sqrt{\kappa}(g^2 \chi_m \lambda_+\lambda_-\sqrt{\kappa}\\
  \notag
  &+\Omega\chi_d\zeta g G^\prime\chi_m \sqrt{\kappa}\lambda_+\lambda_-g\chi_m)x^{in}_a \\
    \label{append1}
\end{align}
 Here, 
\begin{align}
    \zeta= \frac{1}{i\omega +\frac{\Gamma}{2}+ (\Delta_q+2G\bar \chi -2\Omega_R \bar d - 2G^{\prime 2}\chi_m ) \Delta_q\chi_d}
    \label{zeta}
\end{align}

 Also, the modified susceptibility of the electro-mechanical sub-system is given by, $\chi^\prime_d=-\Delta_q \zeta \chi_d$.
To match the susceptibilities of sub-systems, we consider $\Delta_q=\Omega $ then $\chi^\prime_d=-\Omega\zeta\chi_d$. This gets further simplified for, $2G\bar\chi-2\Omega_R \bar d -2 G^{\prime 2}\chi_m=0$. Consequently, the expression \ref{zeta} can be re-written as, \begin{align}
     \zeta=\frac{1}{i\omega +\frac{\Gamma}{2}+\Delta^2_q\chi_d}
 \end{align}

 Hence $\chi^{\prime}_d =\frac{\Omega}{\Omega ^2 - \omega^2 + i \omega \Gamma + \Gamma ^2/4}$. Furthermore, for $g^2\chi^2=1$, \ref{append1} finally takes the form
 
 \begin{align}
   & \sqrt{\kappa}(g^2 \chi_m \lambda_+\lambda_-\sqrt{\kappa}+\Omega\chi_d\zeta g G^\prime\chi_m \sqrt{\kappa}\lambda_+\lambda_-g\chi_m)x^{in}_a \notag \\
    & = \lambda_+ \lambda_-  \kappa (g^2 \chi_m + G^{\prime 2}\chi_d\prime)x^{in}_a
    \label{append2}
\end{align}
This has been used in the expression of $P^{out}_a$ (Eq. \ref{paout}).

 The above mentioned condition, $g^2\chi^2=1$ provides the constraint equation, which can be further simplified as,

 \begin{align}\label{eqqq}
     \frac{g^2\Omega^2}{(\Omega^2-\omega^2)^2+\omega^2\gamma^2_m}=1
 \end{align}

After solving this Eq. (\ref{eqqq}) symbolically, we obtain frequency,
 \begin{equation}
     \omega_{1,2}=\pm\Big[\Omega^2-\gamma^2_m/2-\sqrt{(4\Omega^2g^2-4\Omega\gamma^2_m +\gamma^4_m)/2}  \Big]^{1/2}
 \end{equation}
  \begin{equation}
     \omega_{3,4}=\pm\Big[\Omega^2-\gamma^2_m/2+\sqrt{(4\Omega^2g^2-4\Omega\gamma^2_m +\gamma^4_m)/2}  \Big]^{1/2}
 \end{equation}

When one of the above constraints is satisfied by the parameters $\Omega,~ \gamma,~ g$,  Eq. (\ref{append2}) will hold consequently.

\bibliographystyle{apsrev4-1}

\bibliography{bibb}

\end{document}